\documentclass{aa}
\usepackage{epsfig}
\usepackage{amssymb}

\newcommand{\Pm}{\mbox{\it P\hspace{-0.3pt}m}}
\newcommand{\Rm}{\mbox{\it Rm}}
\newcommand{\Rey}{\mbox{\it Re}}
\newcommand{\bnabla}{\mbox{\boldmath $\nabla$}}
                                        
\newcommand{\beq}{\begin{equation}}
\newcommand{\eeq}{\end{equation}}
\newcommand{\beqarr}{\begin{eqnarray}}
\newcommand{\eeqarr}{\end{eqnarray}}
\newcommand{\barr}{\begin{array}}
\newcommand{\earr}{\end{array}}
\newcommand{\bcent}{\begin{center}}
\newcommand{\ecent}{\end{center}}
\newcommand{\rf}[1]{(\ref{#1})}

\newcommand{\laplace}{\nabla^2}  

\newcommand{\cross}{\wedge}
\newcommand{\grad}{\bnabla}
\newcommand{\dvgnce}{\bnabla \cdot}
\newcommand{\curl}{\bnabla \wedge}

\newcommand{\pder}[3]{\frac{\partial^{#3}#1}{\partial #2^{#3}}}
\newcommand{\pd}[1]{\partial_{#1}}

\newcommand{\bess}{{\mathcal{B}}}

\newcommand{\im}{{\mathrm i}}

\title{Magnetic instability in a sheared azimuthal flow}
\author{A. P. Willis \and C. F. Barenghi}
\institute{Department of Mathematics,  
 University of Newcastle, Newcastle upon Tyne, NE2 7RU, UK}

\date{Received 8 January 2002 / Accepted 4 April 2002}

\begin{document}

\abstract{
We study the magneto-rotational instability of an incompressible
flow which rotates with angular velocity $\Omega(r)=a+b/r^2$ where
$r$ is the radius and $a$ and $b$ are constants.
We find that an applied magnetic field destabilises the flow, in
agreement with the results of \cite{rudiger01}. 
We extend the investigation in the region of parameter space which
is Rayleigh stable.
We also study the instability at values of magnetic Prandtl number
which are much larger and smaller than R\"{u}diger \& Zhang.  
Large magnetic Prandtl numbers are motivated by
their possible relevance in the central region of galaxies
(Kulsrud \& Anderson 1992). In this regime we find that increasing the
magnetic Prandtl number greatly enhances the instability; the stability
boundary drops below the Rayleigh line and tends toward the solid body 
rotation line. Very small magnetic Prandtl numbers are motivated by the
current MHD dynamo experiments performed using liquid sodium and gallium.
Our finding in this regime confirms R\"{u}diger \& Zhang's conjecture
that the linear magneto-rotational instability and the nonlinear
hydrodynamical instability (Richard \& Zahn 1999) take place at Reynolds
numbers of the same order of magnitude.
}

\maketitle

\section {Motivation}

It is thought that turbulence in accretion discs arises from a
magneto-rotational instability (MRI), where a magnetic field destabilises
a rotating velocity field which decreases outwardly. This instability
was discovered by \cite{velikhov59} and \cite{chandrasekhar61} when studying
the motion of a fluid between two concentric cylinders (Taylor-Couette
flow). It was only years later, when it was realized that Couette flow
can be interpreted as model of Keplerian flow, that the implications for
astrophysics were fully appreciated (Balbus \& Hawley 1991).
There have been many studies of the MRI recently, from
numerical simulations of accretion discs (Brandenburg et al. 1995) to
nonlinear calculations in spherical geometry (Drecker et al. 2000).
In particular, in a recent paper \cite{rudiger01}
analysed the linear stability of hydromagnetic Couette flow and showed that,
if the magnetic Prandtl number is less than unity, azimuthal Couette flow is
more easily destabilised with a magnetic field than without. They
found that the instability extends into the region of parameter space
which is Rayleigh-stable without a magnetic field, which is
important since Keplerian rotation is Rayleigh-stable.
R\"{u}diger \& Zhang were also able to study the  MRI instability at magnetic
Prandtl numbers $\Pm$ as small as $0.001$, towards the limit relevant to
liquid sodium  and gallium ($\Pm \approx 10^{-5}$) which are
used in current MHD dynamo experiments (Tilgner 2000).

The aim of this paper to is to extend the investigation of 
\cite{rudiger01}. We explore the instability as a function
of the speed of the outer cylinder in the Rayleigh-stable region,
which is the parameter space of astrophysical interest (R\"{u}diger \& Zhang
considered only one nonzero 
ratio of outer to inner cylinder's rotation).
We also determine the effect of changing the magnetic Prandtl
number, extending the range studied by R\"{u}diger \& Zhang.
       
\section{Model}

We consider an incompressible fluid contained between two
coaxial cylinders of inner radius $R_1$ and outer radius $R_2$ which
rotate at prescribed angular velocities $\Omega_1$ and $\Omega_2$.
We use cylindrical coordinates $(r,\phi,z)$ and assume that
a uniform magnetic field of strength $B_0$ is applied
in the axial $z$ direction.
At small angular
velocities the flow is purely azimuthal (circular-Couette flow) 
and has magnitude
$V_0(r)=ar+b/r$ which corresponds to the rotation law,
\beq
   \label{eq:rotlaw}
   \Omega(r)=\frac{V_0}{r}=a+\frac{b}{r^2},
\eeq
The constants $a$ and $b$ are determined by the no-slip boundary
conditions at the cylinder's walls,
so we have,
\beq
   \label{eq:consts}
   a=\Omega_1 \frac{(R_1^2/R_2^2-\Omega_2/\Omega_1)}{(1-R_1^2/R_2^2)},\quad
   b=\Omega_1 R_1^2 \frac{(1-\Omega_2/\Omega_1)}{(1-R_1^2/R_2^2)}.
\eeq

At sufficiently high angular velocities Couette flow becomes
unstable. The resulting velocity and magnetic fields  
$\vec{V}(r,\phi,z,t)$ and $\vec{B}(r,\phi,z,t)$
are determined by the MHD equations which we write in dimensionless
form as,
\beq
   \label{eq:mom}
   \pder{\vec{V}}{t}{} + (\vec{V} \cdot \grad) \vec{V} =
   - \grad p + \laplace \vec{V} +\frac{Q}{\Pm}(\curl \vec{B})
   \cross \vec{B},
\eeq \beq
   \label{eq:ind} 
   \pder{\vec{B}}{t}{} = \frac1{\Pm} \laplace \vec{B}
   + \curl (\vec{V} \cross \vec{B}),
\eeq \beq
   \dvgnce \vec{B}=0,\quad
   \dvgnce \vec{V}=0,
\eeq
where $p$ is the pressure. In writing \rf{eq:mom} and \rf{eq:ind} we used
$\delta=R_2-R_1$ as unit of length, $\delta^2/\nu$ as unit of time
and $B_0$ as unit of  magnetic field, where $\nu$ is the kinematic
viscosity. The governing dimensionless parameters of the problem
are the Reynolds numbers  $\Rey_1$, $\Rey_2$, radius ratio $\eta$,
and rotation ratio $\mu$,
\beq
   \Rey_1 = \frac{R_1 \Omega_1 \delta}{\nu},\quad
   \Rey_2 = \frac{R_2 \Omega_2 \delta}{\nu},\quad
   \eta = \frac{R_1}{R_2},\quad
   \mu = \frac{\Omega_2}{\Omega_1},
\eeq
together with Chandrasekhar's number $Q$ and the magnetic Prandtl
number $\Pm$,
\beq
   Q=\frac{B_0^2  \delta^2 \sigma}{\rho \nu},\quad
   \Pm=\frac{\nu}{\lambda},
\eeq
where $\rho$ is the density, $\sigma$ is the electrical conductivity,
$\lambda=1/(\sigma \mu_0)$ is the magnetic diffusivity and $\mu_0$ is
the permeability.

We solve the MHD equations by direct time stepping from a small seeding
initial condition and determine the stability of the Couette solution
in different regions of parameter space. 
Our numerical method for 3D nonlinear flow is detailed in 
\cite{willis02}. 
For this linear study it suffices to say that the formulation is
based on representing $\vec{V}$ and $\vec{B}$ with suitable potentials
which are spectrally expanded over Fourier modes 
$\exp[\im m \phi + \im \alpha z]$
in the azimuthal and axial directions and over Chebyshev
polynomials in the radial direction.  
The axial wavelength of the disturbance is $2\pi/\alpha$.
We assume no-slip boundary conditions for $\vec{V}$, and electrically 
insulating boundaries conditions for $\vec{B}$,
\beq
   \pder{B_r}{z}{} = \frac{B_z}{\bess_m}\,\pder{\bess_m}{r}{},
   \quad
   \frac1{r}\,\pder{B_z}{\phi}{} = \pder{B_\phi}{z}{},
\eeq 
where $\bess_m(r)$ denotes the modified Bessel functions, $I_m(\alpha r)$ 
on $R_1$, and $K_m(\alpha r)$ on $R_2$.
The time stepping is based on a
combination of second order accurate Crank-Nicolson and Adams-Bashforth
methods. The resulting computer code was tested against published
results with and without magnetic field (Chandrasekhar 1961;
Roberts 1964; Marcus 1984; Jones 1985; Barenghi 1991).

\section{Results}

It is well known that, in the absence of magnetic field, inviscid
Couette flow is linearly stable provided that the celebrated Rayleigh
criterion ($\mu>\eta^2$) 
is satisfied.  

\begin{figure}
   \epsfig{figure=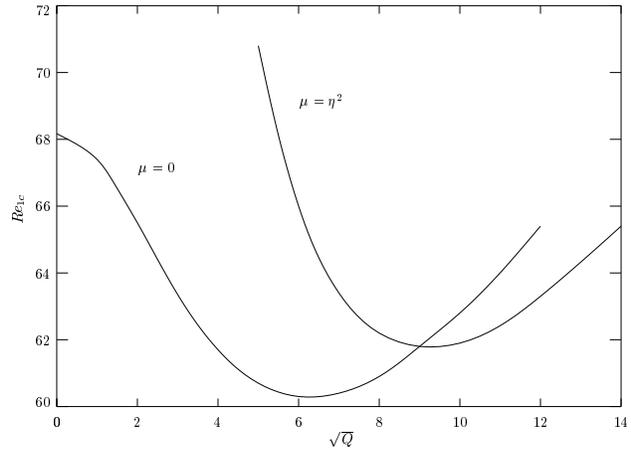, scale=0.56}
   \caption{
   $\eta=0.5$, $\Pm=1$.  Critical Reynolds number of the
   inner cylinder ($\Rey_{1c}$) versus applied magnetic
   field ($\sqrt{Q}$); two cases are presented: outer cylinder fixed
   ($\mu=0$) and on the Rayleigh line ($\mu=\eta^2$).
} \end{figure}
Figure 1 
shows the result of our calculations for
the stability of dissipative Couette flow to axisymmetric perturbations,
for radius ratio
$\eta=0.5$ and magnetic Prandtl number $\Pm=1$, as a function of the 
applied magnetic field. We plot the result in terms of $\sqrt{Q}$ rather
than $Q$ in order to make direct comparison with the work of 
\cite{rudiger01}.  The first curve refers to the case in which the
outer cylinder is fixed, $\mu=0$, which is the most studied case in the
fluid dynamics literature. It is apparent that the presence of
a magnetic field makes the flow more unstable.  
The critical Reynolds number, which is $\Rey_{1c}=68.2$ for $Q=0$,
decreases with increasing $Q$ and has a minimum at $Q=39$. The most
unstable mode is $m=0$ over the range for $Q$ in Fig. 1.  From here we
consider axisymmetric disturbances only.  The critical axial wavenumber 
$\alpha$ decreases significantly from $3.1$ to $1.7$
over the range, and varies like $1/\sqrt{Q}$ thereafter.  This stiffening
eventually restabilises the flow and $\Rey_{1c}$ increases like $\sqrt{Q}$
for strong fields.

The initial destabilisation is consistent with the finding of
\cite{rudiger01};  the small difference between their $\Rey_{1c}$
and ours is certainly due to the different boundary
conditions for $\vec{B}$ (they assumed pseudo-vacuum conditions
and we assume insulating conditions). The second curve of Fig. 1 refers
to the case $\mu=\eta^2$ (the Rayleigh line), which separates
stable and unstable regions
in the absence of a magnetic field. The curve well illustrates the striking
destabilising effect of the magnetic field.

\begin{figure}
   \epsfig{figure=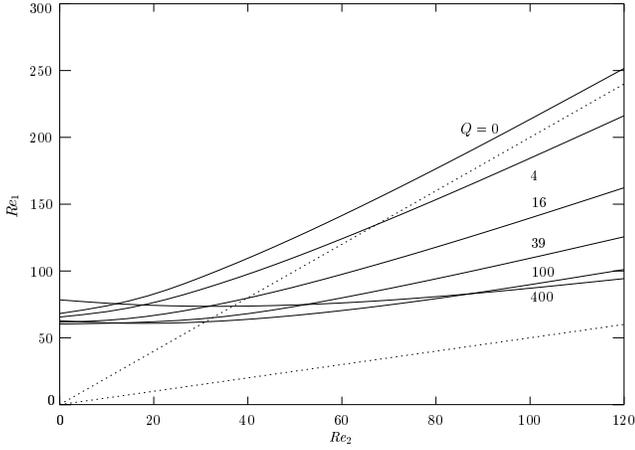, scale=0.56}
   \caption{
   $\eta=0.5$, $\Pm=1$.  Critical Reynolds number of the
   inner cylinder ($\Rey_{1c}$) versus Reynolds number of the outer cylinder
   ($\Rey_2$) at different values of applied magnetic field ($Q$).
   The upper dotted line is the Rayleigh criterion.
   The lower dotted line is solid-body rotation.}
\end{figure}
Figure 2 
shows stability boundaries in the $\Rey_1$ vs $\Rey_2$ plane for
different values of the imposed magnetic field.  The axial wavenumber 
$\alpha$ does not vary a great deal along the boundaries, its dependence
being principally determined by the strength of the field $Q$.

It is apparent that
even a small value of $Q$ is enough to make the boundary cross the
Rayleigh line (the upper dotted curve in the figure). The destabilising
effect of the magnetic field is so large that the stability boundary drops
toward the region of solid body rotation ($\Omega_1=\Omega_2$ or $\mu=1$),
which is the lower dotted line.
However, if the applied field is strong enough the flow can be
restabilised, in accordance with the results of Fig. 1.

\begin{figure}
   \epsfig{figure=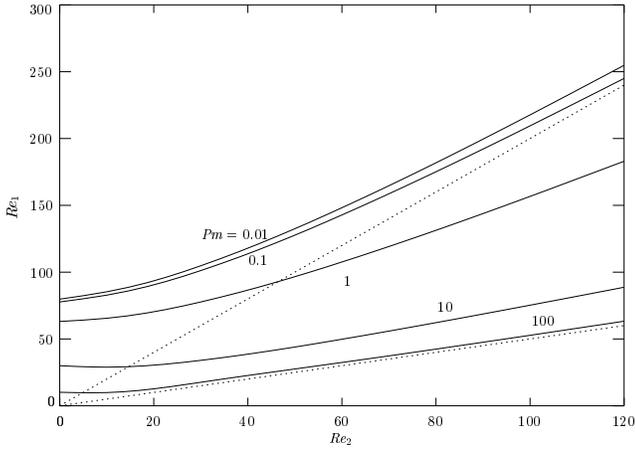, scale=0.56}
   \caption{
   $\eta=0.5$, $Q=10$.  Critical Reynolds number of the
   inner cylinder ($\Rey_{1c}$) versus Reynolds number of the outer cylinder
   ($\Rey_2$) at different values of magnetic Prandtl number ($\Pm$).
   The upper dotted line is the Rayleigh criterion.
   The lower dotted line is solid-body rotation.}
\end{figure}
The destabilisation becomes more important the larger the magnetic
Prandtl number $\Pm$.  Figure 3 
shows results for a moderate value
of applied magnetic field, $Q=10$, at increasing values of $\Pm$. It is
apparent that the stability boundary drops much below the Rayleigh line
$\mu=\eta^2$ and, for large enough $\Pm$, becomes asymptotic to the 
solid body rotation line $\mu=1$.

\begin{figure}
   \epsfig{figure=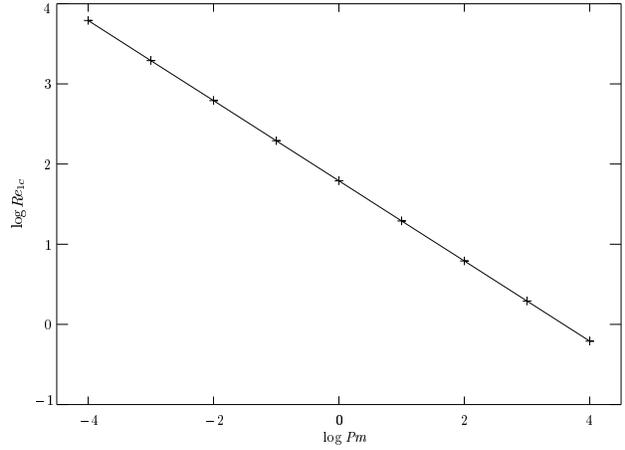, scale=0.56}
   \caption{
   $\eta=0.5$.  Critical Reynolds number of the
   inner cylinder ($\Rey_{1c}$) versus magnetic Prandtl number ($\Pm$)
   on the Rayleigh line, $\mu=\eta^2$.
} \end{figure}
Finally, in Fig. 4 we find 
that on the Rayleigh line the stability
follows a power law,
\beq
   \label{eq:Rey-Pm_num}
   \Rey_{1c} \propto \Pm^{-\beta}, \quad
   \beta = -0.500\pm 0.002 ,
\eeq
over a surprising range of $\Pm$.
The error in $\beta$ is based on a fit through all points
calculated.
For $\eta=0.5$ the minimum $\Rey_{1c}$ occurs at 
$Q=86$ and $\alpha=2.1$.  
To verify this result we use the WKB analysis of \cite{ji01}.
The local dispersion relation they find for the Taylor-Couette flow 
is identical to the one derived for accretion discs in the incompressible 
limit (Sano \& Miyama 1999).
From the dispersion relation the necessary and sufficient
condition for stability to axisymmetric perturbations is,
\beq
   \label{eq:dispersion}
   (\Pm+S^2)^2(1+\epsilon^2) + 2\zeta\Rm^2 - 2(2-\zeta)\Rm^2 S^2 \ge 0 ,
\eeq
where $S$ is the ratio of diffusive to Alfv\'en timescales, 
$\Rm$ is the magnetic Reynolds number, 
and $\zeta = 1/(r\Omega)\,\pd{r}(r^2\Omega)$.
On the Rayleigh line, from 
\rf{eq:rotlaw} and \rf{eq:consts},
$\Omega(r)$ 
takes the simple form,
$\Omega = b/r^2$,
and so $\zeta=0$ at all $r$.
We said earlier that in Fig. 2 the wavenumber changes little along
the boundary, as its dependence is principally on the strength of the
imposed field.  
The wavenumber parameter $\epsilon$ of \cite{ji01} 
then depends only on $Q$, so 
$\epsilon = \epsilon(Q)$.  
Ignoring factors that depend only on $\epsilon$ in
the substitutions,
\beq
   S^2 = \Pm \, Q,  \quad  \Rm = \Pm \, \Rey,
\eeq
\rf{eq:dispersion} leads to the condition for stability,
\beq
   \label{eq:Rey-Pm}
   \Rey_1 \le \frac{A(Q)}{\sqrt{\Pm}},
\eeq
which confirms our result \rf{eq:Rey-Pm_num}.
This relation \rf{eq:Rey-Pm} should hold for all $\Pm$ as we have not 
been required to take any limits.  
The function $A(Q)$ has a minimum for some $Q$ and $\alpha(Q)$.
For $\eta=0.5$ we found the minimum occurs at $Q=86$, $\alpha=2.1$.

Whilst the linear result \rf{eq:Rey-Pm} should hold for very small $\Pm$,
from the work of Wendt (1933) and Taylor (1936) a turbulent
(nonlinear) instability
is observed at large Reynolds numbers ($\Rey_1\gtrsim 2.5\times 10^5$)
with this radius ratio.
Extrapolating down one magnitude
to $\Pm=10^{-5}$ for laboratory fluids gives
$\Rey_{1c}\approx 2\times 10^5$, a similar value,
 for the magneto-rotational (linear) instability.  

A radial truncation of 12 Chebyshev modes was found sufficient for all 
calculations.  In particular, for the calculation at $\Pm=10^{-4}$ 
in Fig. 4, convergence was tested by increasing the truncation and also
decreasing the timestep.
The  fractional numerical error in 
$\Rey_{1c}$ is estimated at approximately $10^{-6}$.

\section{Discussion}

Our calculations show that many rotation laws of the form
$\Omega(r)=a+b/r^2$ which are hydrodynamically stable (that is
to say, they satisfy the Rayleigh criterion) become linearly unstable
when a magnetic field is applied. Our results confirm the
finding of \cite{rudiger01} and extend them in the Rayleigh
stable region. 

We have determined the instability at magnetic Prandtl
numbers $\Pm$ one order of magnitude smaller than R\"{u}diger \& Zhang's,
towards the small magnetic Prandtl
number limit, which is relevant to possible MHD dynamo experiments with
liquid sodium and gallium.
Although the power law $\Rey_{1c}\propto\Pm^{-0.5}$ that we find 
on the Rayleigh line ($\mu=\eta^2$)
is slightly different from theirs ($\Rey_{1c}\propto\Pm^{-0.65}$ on 
$\mu=\frac1{3}$), it confirms 
their conjecture that the nonlinear instability found by 
\cite{richards99} and the MRI are likely to occur at Reynolds
numbers of the same order of magnitude.

We also find that the flow becomes particularly unstable if the
magnetic Prandtl number is greater than unity. 
The instability boundary in the $\Rey_1$ vs $\Rey_2$
plane rapidly tends towards the solid body rotation line.
This enhanced instability for large $\Pm$ is consistent with 
earlier results of \cite{kurzweg63}.
His boundary conditions were selected such as to avoid mathematical 
difficulties but for small $\Pm$ agreed well with the 
results of \cite{chandrasekhar61}.
The significance of the instability in this case is linked to
the possibility (Kulsrud \& Anderson 1992; Brandenburg 2001)
that large values of $\Pm$ exist in central regions of galaxies.

\end{document}